\documentclass[tighten, twocolumn, twocolappendix]{aastex631}

\usepackage{amsfonts} 
\usepackage{amsmath}
\usepackage{amssymb}
\usepackage{amsthm}
\usepackage{braket}
\usepackage{bm}
\usepackage{cancel}
\usepackage{color}
\usepackage{cuted}
\usepackage{diagbox}
\usepackage{epsfig}
\usepackage{esint}
\usepackage{graphicx}
\usepackage{mathrsfs}
\usepackage{mathtools}
\usepackage{slashed}
\usepackage{subfigure}
\usepackage{xcolor}

\hypersetup{linkcolor=blue, citecolor=magenta, filecolor=cyan, urlcolor=magenta}

\newcommand{\RV}[1]{\textcolor{black}{#1}}

\begin{document}

\title{Toward General-Relativistic Magnetohydrodynamics Simulations\\
in Stationary Non-Vacuum Spacetimes}

\author[0000-0001-7386-7439]{Prashant Kocherlakota}
\affiliation{Black Hole Initiative at Harvard University, 20 Garden Street, Cambridge MA 02138, USA}
\affiliation{Harvard-Smithsonian Center for Astrophysics, 60 Garden Street, Cambridge MA 02138, USA}

\author[0000-0002-1919-2730]{Ramesh Narayan}
\affiliation{Harvard-Smithsonian Center for Astrophysics, 60 Garden Street, Cambridge MA 02138, USA}
\affiliation{Black Hole Initiative at Harvard University, 20 Garden Street, Cambridge MA 02138, USA}

\author[0000-0002-2825-3590]{Koushik Chatterjee}
\affiliation{Black Hole Initiative at Harvard University, 20 Garden Street, Cambridge MA 02138, USA}
\affiliation{Harvard-Smithsonian Center for Astrophysics, 60 Garden Street, Cambridge MA 02138, USA}

\author[0000-0002-3945-6342]{Alejandro Cruz-Osorio}
\affiliation{Institut f{\"u}r Theoretische Physik, Goethe Universit{\"a}t, Max-von-Laue-Str. 1, 60438 Frankfurt am Main, Germany}

\author[0000-0002-8131-6730]{Yosuke Mizuno}
\affiliation{Tsung-Dao Lee Institute, Shanghai Jiao-Tong University, Shanghai, 520 Shengrong Road, 201210, China}
\affiliation{School of Physics \& Astronomy, Shanghai Jiao-Tong University, Shanghai, 800 Dongchuan Road, 200240, China}
\affiliation{Institut f{\"u}r Theoretische Physik, Goethe Universit{\"a}t, Max-von-Laue-Str. 1, 60438 Frankfurt am Main, Germany}

\begin{abstract}

Accretion of magnetized gas on compact astrophysical objects such as black holes has been successfully modeled using general relativistic magnetohydrodynamic (GRMHD) simulations. These simulations have largely been performed in the Kerr metric, which describes the spacetime of a vacuum and stationary spinning black hole (BH) in general relativity (GR). The simulations have revealed important clues on the physics of accretion \RV{flows} and jets near the BH event horizon, and have been used to interpret recent Event Horizon Telescope images of the supermassive BHs, M87$^*$ and Sgr A$^*$. GRMHD simulations require the spacetime metric \RV{to be given} in horizon-penetrating coordinates such that all metric coefficients are regular at the event horizon. \RV{Only a few metrics, notably the} Kerr metric and its electrically charged spinning analog, the Kerr-Newman metric, are currently available in such coordinates. We report here horizon-penetrating forms of a large class of stationary, axisymmetric, spinning metrics. These can be used to carry out GRMHD simulations of accretion on spinning, nonvacuum BHs and non-BHs within GR, as well as accretion on spinning objects described by non-GR metric theories of gravity.
\end{abstract}


\section{Introduction} 
\label{sec:Introduction}

General relativistic magneto-hydrodynamic (GRMHD) simulations have emerged as an indispensable tool in modern astrophysical research, providing a robust framework for investigating the complex dynamics of magnetized gas within extreme gravitational environments \citep[e.g.,][]{Banyuls+1997,Koide+1999, Gammie+2003, DeVilliers+2003}. By virtue of their ability to simulate diverse accretion regimes, spanning from sub-to-super Eddington rates, GRMHD simulations offer invaluable insights into the structure and evolution of a variety of black hole (BH) and neutron star systems \citep[e.g.,][]{Fragile+2007, Narayan+2012,Sadowski+2015,Parfrey+2017, Liska+2018, Porth+2019, Begelman+2022, Chatterjee+2022}, as well as astrophysical phenomena like gamma-ray bursts \citep[e.g.,][]{Gottlieb+2022}, tidal disruption events \citep[e.g.,][]{Curd+2019, Andalman+2022} and high energy flares from the centers of galaxies \citep[e.g.,][]{Chatterjee+2021, Porth+2021, Ripperda+2022}. Furthermore, GRMHD simulations self-consistently launch powerful jets that can accelerate to relativistic speeds and extend to galactic scales, and thus are crucial for understanding galaxy evolution \citep[e.g.,][]{McKinney2006, Tchekhovskoy+2011, Chatterjee+2019, Narayan+2022, Ricarte+2023}. Also, the predictive capabilities of GRMHD simulations facilitate direct comparisons with \RV{observational} data, enhancing their applicability in interpreting real astrophysical systems \citep[e.g.,][]{Moscibrodzka+2009, Dexter+2012, Davelaar+2018, Chael+2019, Chatterjee+2020, Ricarte+2020, Cruz-Osorio+2022, Ressler+2023}, and have helped motivate experimental tests of gravitational effects such as frame-dragging \citep{Ricarte+2022} and electromagnetic extraction of energy from spinning BHs \citep{Chael+2023}.

The Event Horizon Telescope (EHT) collaboration recently demonstrated that it is now possible to ``measure'' the spacetime metric of astrophysical BHs. More concretely, we are able to measure the level of agreement of the spacetime geometry of the supermassive BHs, M87$^*$ \citep{EHTC+2019e, EHTC+2019f, Psaltis+2020} and Sgr A$^*$ \citep{EHTC+2022e, EHTC+2022f}, with that of a Kerr BH \citep{Kerr1963}, \RV{which is} a vacuum, stationary solution of general relativity (GR; see, e.g., \citealt{Wiltshire+2009}). Such investigations allow us, in principle, to test fundamental aspects of GR such as the no-hair hypothesis (see, e.g., \citealt{Carter1971}), and whether astrophysical BHs truly have no surfaces and are devoid of matter \citep{EHTC+2022f}. 

It has also been possible to test the level of agreement of EHT data with non-BH spacetime geometries (such as boson stars, wormholes, classical naked singularities, etc.) within GR, as well as both BH and non-BH spacetimes in non-GR theories \citep{Volkel+2021, Kocherlakota+2021, EHTC+2022f, Vagnozzi+2022}. This program has seen intense activity, and current approaches use a combination of GRMHD simulations in Kerr BH spacetimes (see, e.g., \citealt{EHTC+2019f, EHTC+2022f}) as well as non-GR MHD (``xGRMHD'') simulations in non-rotating spacetimes \citep{Mizuno+2018, Olivares+2020, Fromm+2021, Roder+2023}. In addition, a diverse range of instructive semi-analytic accretion flow models have been considered (see, e.g., \citealt{Broderick+2014, Shaikh+2019a, Shaikh+2019b, Narayan+2019, Paul+2020, Bauer+2022, Ozel+2022, Younsi+2023, Kocherlakota+2022, Ayzenberg2022, EHTC+2022f}) to model the physics of accretion flows and their emission. 

\RV{Nearly} all the xGRMHD work to date has been restricted to spherically symmetric ``static" spacetime metrics. \RV{This is because, apart from the spinning Kerr metric \citep{Kerr1963} and Kerr-Newman metric \citep{Newman+1965b}}, the only metrics available in the horizon-penetrating ``Kerr-Schild" coordinates required by current GRMHD codes (see, e.g., \citealt{Font+1998, McKinney+2004, Sadowski+2014, Porth+2017, Liska+2022} \RV{for a discussion of representative GRMHD codes) are static metrics. The one other exception is  \citet{Nampalliwar+2022}, who use the Kerr-Schild form of the  \cite{Johannsen2013b} stationary metric, which is derived from the well-known \citet{Johannsen+2011} parametrized BH metric.}

Astrophysical objects are expected to possess angular momentum. Hence their spacetimes will in general be axisymmetric (not spherically symmetric), and their metrics should be ``stationary" (not static). For a more realistic confrontation of the underlying effective theory of gravity and fields in the vicinity of ultracompact objects to current observations, and in anticipation of future higher-quality data, it is necessary to be able to model accretion and emission processes in generic stationary spacetimes. However, other than the \RV{aforementioned Kerr, Kerr-Newman and  Johannsen metrics}, no other stationary spacetime metrics have been written down in the horizon-penetrating coordinates needed for computer simulations. The goal of the present paper is to eliminate this roadblock.

We introduce a general family of stationary and axisymmetric metrics \citep{Azreg-Ainou2014a, Azreg-Ainou2014b} which are derived from a corresponding family of ``seed" spherically symmetric metrics. We then cast these stationary metrics in horizon-penetrating ``Kerr-Schild'' form. This is the main contribution of the present paper. Using the results described in this paper, we will report elsewhere the first high-resolution, 3D GRMHD simulations in spinning non-Kerr spacetimes \citep{Chatterjee+2023a, Chatterjee+2023b}. 

We begin in section \ref{sec:Sec2_Static_Metric} by introducing the general form of a (spherically-symmetric) static metric. This form includes the Schwarzschild metric as a special case but covers a wide range of other models as well. We pick out and discuss a few well-known examples of the latter. 

In section \ref{sec:Sec3_AA_Metric} we introduce the Azreg-A{\"i}nou (AA; \citealt{Azreg-Ainou2014b}) metric, which is a stationary, axisymmetric generalization of the static metric described in section \ref{sec:Sec2_Static_Metric}. \RV{The AA metric was proposed as an \textit{ansatz}, inspired by the \cite{Newman+1965a} algorithm, to describe generic spinning spacetimes. This metric has several attractive features, which we touch on in Sec. \ref{sec:Background_AA} and consider in greater detail in Sec. \ref{sec:Sec3_AA_Metric}.}

In section \ref{sec:Sec4_siKS_Form_AA_Metric} we report the form of the AA metric in horizon-penetrating coordinates, as needed for most GRMHD codes, and \RV{in Sec. \ref{sec:Sec4.1_3+1_Form} we write down} the $3+1$ decomposition of the metric required for certain \RV{other} codes. The spherical-polar Kerr-Schild coordinates we use to write the horizon-penetrating AA metric are adapted to the ingoing principal null congruence of spacetime. 

Finally, in section \ref{sec:Sec5_Conclusions} we summarize our findings and present a representative list of stationary and axisymmetric metrics that are obtained from the AA metric for specific choices of its metric functions. These metrics may be directly used in future GRMHD simulations. We also describe the steps needed to generate other stationary metrics in the future. Throughout we use geometrized units, $G = c = 1$, and the metric signature is $(-, +, +, +)$.

\RV{Of necessity, this paper includes a fair amount of technical discussion about the AA metric and its various forms. A reader who is interested in utilizing various BH and non-BH metrics to explore with GRMHD simulations, can directly obtain the main results by reading only sections \ref{sec:Sec2_Static_Metric} and \ref{sec:Sec5_Conclusions}, along with tables \ref{table:Known_Static_Solutions} and \ref{table:siKS_Known_Solutions}.}

\subsection{\RV{Background on the Azreg-A{\"i}nou Metric}}
\label{sec:Background_AA}
\RV{While in general the description of an arbitrary stationary and axisymmetric spacetime requires ten independent metric functions, the AA metric remarkably makes use of only three functions. Furthermore, when considered as a generalization of a static metric, two of these functions are fixed by the original static metric functions. The one remaining free function is a conformal factor that can be fixed by solving the appropriate field equations. Moreover, as we discuss, there is a naturally attractive choice for this third function such that it is no longer a free function. Thus, the AA metric is both a powerful and surprisingly simple \textit{ansatz} for spinning spacetimes.}

\RV{These simplifications are partly a consequence of the AA metric describing only circular spacetimes, a particularly important subfamily of stationary and axisymmetric spacetimes which require at most five free functions (see, e.g., Sec. 2.2 of \citealt{Gourgoulhon2010}). The further reduction to just three metric functions (or indeed only two functions in the most natural from of the AA metric) is achieved via a novel modification in \cite{Azreg-Ainou2014b} to the original Newman-Janis algorithm, where the regularity of a coordinate transformation is used to eliminate the \textit{ad hoc} ``decomplexification step'' that was required in \cite{Newman+1965a} (cf. also \citealt{Rajan2016}).}

\RV{Circular spacetimes have been of central interest in explorations of non-Kerr spacetimes, and several popular parametrized metrics have been constructed to characterize and study their varied properties. The Johannsen-Psaltis (JP; \citealt{Johannsen+2011, Johannsen2013b}) framework has had success, e.g., in establishing tests of the no-hair conjecture \citep{Johannsen+2010} and of the post-Newtonian structure of astrophysical BH spacetimes \citep{Psaltis+2020}. The Konoplya-Rezzolla-Zhidenko (KRZ; \citealt{Rezzolla+2014, Konoplya+2016}) framework has been used, e.g., to map well-known static metrics onto a single low-dimensional space with high accuracy to enable comparisons between spacetimes \citep{Kocherlakota+2020} and to test BH metrics with gravitational-wave as well as X-ray observations \citep{Volkel+2020, Cardenas-Avendano+2020}. The current \citep{Volkel+2021, Kocherlakota+2022, Younsi+2023} and future \citep{Ayzenberg2022, Kocherlakota+2023} ability to use BH images to test gravity has also been demonstrated using both these frameworks.}

\RV{We note that while the JP and KRZ parametric metrics allow detailed explorations of black hole spacetimes, the AA \textit{ansatz} metric can characterize also non-BH objects such as naked singularities, wormholes, boson stars, etc. Furthermore, it is clear to see from \cite{Johannsen2013b} that the JP metric uses a fundamentally different \textit{ansatz} compared to the AA metric to describe circular spacetimes, requiring four metric functions (see eq. 51 there), and motivated by imposing that geodesics be Lioville-separable (eq. 10 there), i.e., that they possess a Carter constant. This is similarly true for the KRZ metric, where five free functions are permitted (see eq. 7 there). \cite{Konoplya+2021} explore a subclass of the KRZ metric that admits a Carter constant. As we show later, null geodesics in the AA metric always possess a Carter constant, and in the simplest form of the AA metric (which we favor) time-like geodesics also have a Carter constant. We note that several other parametrization frameworks have been constructed and used successfully to investigate observable effects due to modifications of the spacetime metric in the strong-field regime (see, e.g., \citealt{Vigeland+2011, Carson+2020}). For the status of parameterizations of non-circular spacetimes, we direct the reader to \cite{Delaporte+2022}.}

\RV{In addition to its simplicity and appealing properties discussed above}, our choice to use the class of AA metrics is further motivated by the knowledge that nearly all well-known solution metrics across various theories of gravity can be cast in this form. This metric, therefore, presents an excellent starting point for a broad forward-modeling study of the effect of the spacetime metric on various observables of interest. The Liouville-separability of the null geodesic equations makes computing various characteristic features of the spacetime such as the location of the photon shell or the shape of the shadow boundary (see, e.g., \citealt{Shaikh2019}, \citealt{Kocherlakota+2021}, \citealt{Solanki+2022}) analytically tractable. Similarly, the separability of the timelike orbits facilitates a study of, e.g., equatorial Keplerian orbits, and allows us to go one step further with semi-analytic techniques. Details regarding geodesic orbits can be found in appendix \ref{sec:AppC_Separability_AA_Metric}. 

\section{General Class of Static Metrics}
\label{sec:Sec2_Static_Metric}

\begin{table*}
\begin{center}
\caption{\textit{Metric functions of selected spherically-symmetric and static spacetimes.} 
We have used the coordinate freedom to set $-\hat{g}_{tt}\hat{g}_{rr}=g(r)=1$ in the general metric \eqref{eq:Static_Spherical_Seed_Metric}, hence we have only two metric functions, $f(r)$ and $R(r)$. The Schwarzschild and Reissner-Nordstr{\"o}m (RN) spacetimes describe the vacuum and the electrovacuum black hole (BH) solutions of general relativity (GR), respectively. The Modified Hayward BH spacetime contains a regular BH and is generated by an anisotropic fluid. The GMGHS BH metric is a solution to the low energy theory of the heterotic string, and contains a scalar field (a dilaton, which vanishes asymptotically) as well as an electromagnetic field. The JNW naked singularity spacetime is generated by a scalar field. The JMN-1 naked singularity spacetime is an interior solution to the Schwarzschild spacetime and is generated by an anisotropic fluid. For easier comparison with the other five models, the JMN-1 metric is presented here with $g(r) = 1$ (see Appendix \ref{sec:AppA_g1_Form} for details), thus differing from the original form given in eq.~(29) of \cite{Joshi+2011}. The parameter $Q$ denotes an electromagnetic charge, $L$ denotes a (de Sitter) length scale, $R_b \geq 2.5M$ denotes a matching radius ($\sigma = M/(R_{\mathrm{b}}-M)$ and $r_{\mathrm{b}} = (1-\sigma)R_{\mathrm{b}}$), and $\hat{\nu}$ denotes a scalar charge ($r_\star=2M/(1-\hat{\nu})$ locates the curvature singularity).}
\label{table:Known_Static_Solutions}
\renewcommand{\arraystretch}{1.5}
\centering
\begin{tabular}[t]{|l|l|l|c|c|}
\hline
Object & Theory & Spacetime & $f(r)$ & $R^2(r)$ \\
\hline
Black Hole & GR & Schwarzschild & $1-\frac{2M}{r}$ & $r^2$ \\ 
Black Hole & GR & RN  & $1 - \frac{2M}{r} + \frac{Q^2}{r^2}$ & $r^2$ \\ 
Black Hole & GR & Modified Hayward & $1 - \frac{2Mr^3}{r^4+2L^4}$ & $r^2$ \\ 
Black Hole & String & GMGHS & $1- \frac{2M}{r}$ & $r\left(r-\frac{Q^2}{M}\right)$ \\
Naked Singularity & GR & JNW & $\left(1-\frac{r_\star}{r}\right)^{1-\hat{\nu}}$ & $r^2\left(1-\frac{r_\star}{r}\right)^{\hat{\nu}}$ \\
Naked Singularity & GR & JMN-1 & $\left(1-\frac{2M}{R_{\mathrm{b}}}\right)\left(\frac{r}{r_{\mathrm{b}}}\right)^{2\sigma}$ & $R_{\mathrm{b}}^2\left(\frac{r}{r_{\mathrm{b}}}\right)^{2-2\sigma}$ \\
\hline
\end{tabular}
\end{center}
\end{table*}

The line element $\mathrm{d}s^2 = \hat{g}_{\mu\nu}\mathrm{d}x^\mu\mathrm{d}x^\nu$ of an arbitrary static and spherically-symmetric metric $\hat{g}_{\mu\nu}$ can be written in spherical-polar coordinates, $x^\mu = (t, r, \vartheta, \varphi)$, in the form 
\begin{equation} \label{eq:Static_Spherical_Seed_Metric}
\mathrm{d}s^2 =\ -f(r)\mathrm{d}t^2 + \frac{g(r)}{f(r)}~\mathrm{d}r^2 + R^2(r)~\mathrm{d}\Omega_2^2\,,
\end{equation}
where $\mathrm{d}\Omega_2^2 = \mathrm{d}\vartheta^2 + \sin^2{\vartheta}~\mathrm{d}\varphi^2$ is the standard line-element on a unit 2-sphere. This form of the metric involves three functions of the coordinate $r$: $f(r)$, $g(r)$, $R(r)$. However, an arbitrary static metric can be defined using only two metric functions, i.e., any one of the metric functions, $f$, $g$, $R$, can be eliminated by an appropriate change of coordinates (see, e.g., Ch. 14 of \citealt{Plebanski+2012}). For example, it is possible to find a different radial coordinate $r$ in which $-\hat{g}_{tt}\hat{g}_{rr} = g(r) = 1$ (as we do in Table~\ref{table:Known_Static_Solutions}, and discuss further in Appendix \ref{sec:AppA_g1_Form}). However, the required coordinate transformation does not always lead to analytically convenient metric functions. We therefore keep \RV{an arbitrary} $g(r) \RV{> 0}$ in \eqref{eq:Static_Spherical_Seed_Metric}.

The function $R(r)$ gives the curvature-radius or areal-radius as a function of the radial coordinate $r$. In simple models, $R(r)=r$ (see Table \ref{table:Known_Static_Solutions}). When this is the case, the coordinate-radius $r$ of a 2-sphere is its curvature-radius, $\kappa_{\mathrm{G}}(r) = \mathscr{R}^{(2)}(r)/2 = 1/r^2$, as well as its areal-radius, $\mathcal{A}(r) = 4\pi r^2$. Here we have used $\kappa_{\mathrm{G}}$, $\mathscr{R}^{(2)}$ and $\mathcal{A}$ to denote the Gaussian curvature, the Ricci scalar, and the area of a 2-sphere, respectively. While it is always possible to find a coordinate transformation to curvature coordinates in which $R(r)=r$, \RV{this again} may only be tractable numerically for some models. Finding simple analytic forms for all the metric functions in such coordinates is not in general possible (see, e.g., Sec. IV D of \citealt{Kocherlakota+2020}). Thus, for general convenience, we will leave $R(r)$ free. 

When the metric \eqref{eq:Static_Spherical_Seed_Metric} describes a BH spacetime, its event horizon is a null, stationary surface. The normal to such a constant-$r$ surface satisfies $\hat{g}^{\mu\nu}\partial_\mu r \cdot \partial_\nu r = 0$, from which we find that the horizon is located at the outermost root of $\hat{g}^{rr}=0$, i.e., at the outermost root of $f(r)$.

In Table \ref{table:Known_Static_Solutions} we list the metric functions corresponding to six well-known and representative spherically-symmetric static solutions in GR and in string theory. For uniformity, we have written all the metrics in coordinates such that $g(r)=1$. The first two solutions listed are the canonical nonrotating BH solutions of GR, namely, the Schwarzschild BH and the charged Reissner-Nordstr{\"o}m (RN) BH, which contain, respectively, a spacelike and a timelike curvature singularity (see, e.g., \citealt{Poisson2004}). Several regular BH spacetime models have been proposed to mimic the desired effect of singularity-resolving physics (see, e.g., \citealt{Bardeen1968, Ayon-Beato+1998, Bronnikov2001, Dymnikova2004, Hayward2006}). The Modified Hayward BH (\citealt{Zhou+2023a}; $n=4$) spacetime included in Table~\ref{table:Known_Static_Solutions} is generated by an anisotropic fluid with a radial equation of state, $\omega_r = p_r/\rho = -1$, where $\rho$ is the total energy density and $p_r$ is the radial pressure in the fluid rest-frame. The fluid is dark-energy-like, producing the pressure necessary to ``avoid'' a singularity. The charged Gibbons-Maeda-Garfinkle-Horowitz-Strominger (GMGHS) BH \citep{Gibbons+1988, Garfinkle+1991} is different from the RN BH of GR because of the presence of a dilaton (a scalar field) that mediates the interaction between electromagnetism and gravity in the low-energy effective action of string theory. In particular, the central singularity of the GMGHS BH remains spacelike. 

In addition to the above four BH solutions, Table~\ref{table:Known_Static_Solutions} lists two other non-BH models. Over the years, non-BH solutions to various theories have been discussed in the literature and significant progress has been made to determine ways in which such objects can be distinguished from BHs using observations (see, e.g., \citealt{Kocherlakota+2021, EHTC+2022f}). The Janis-Newman-Winicour naked singularity (JNW; \citealt{Janis+1968}) spacetime is a static solution generated by a minimally-coupled, massless scalar field in GR. Indeed, this is also a solution to string theory when an electromagnetic field is absent \citep{Virbhadra1997} and is in addition a solution to the Brans-Dicke theory with the parameter $\omega = -1$ \citep{Kar1997}. Finally, the Joshi-Malafarina-Narayan-1 naked singularity (JMN-1; \citealt{Joshi+2011}) is constructed in GR using an anisotropic fluid with vanishing radial pressure, $p_r = 0$.

\section{Generalization to Stationary Metrics}
\label{sec:Sec3_AA_Metric}

Inspired by the \citet{Newman+1965a} algorithm, \cite{Azreg-Ainou2014b} proposed a stationary and axisymmetric generalization of the spherically-symmetric metric \eqref{eq:Static_Spherical_Seed_Metric} described in section \ref{sec:Sec2_Static_Metric}. We will refer to this stationary metric as the AA metric and denote it by $g_{\mu\nu}$ (to distinguish it from $\hat{g}_{\mu\nu}$ of the spherically symmetric metric). In Boyer-Lindquist (BL; \citealt{Boyer+1967}) coordinates, $x^\mu = (t, r, \vartheta, \varphi)$,%
\footnote{Note that we do not distinguish between the labels for the radial coordinates used to write the non-spinning \eqref{eq:Static_Spherical_Seed_Metric} and the spinning \eqref{eq:AA_Metric} metrics to avoid a proliferation of symbols. Furthermore, while the metric signature here $(-,+,+,+)$ differs from that in \cite{Azreg-Ainou2014b}, it can be checked that eq.~\ref{eq:AA_Metric} corresponds to eq. 16 there.} the AA metric is given by
\begin{align} \label{eq:AA_Metric}
\mathrm{d}s^2
=&\ \frac{X}{\Sigma}\left[
-\left(1-\frac{2 F}{\Sigma}\right)\mathrm{d}t^2 
-2\frac{2 F}{\Sigma}a\sin^2{\vartheta}~\mathrm{d}t\mathrm{d}\mathrm{\varphi}
\right. \nonumber \\
&\ \left. 
+ \frac{\Sigma}{\Delta}\mathrm{d}r^2 + \Sigma~\mathrm{d}\vartheta^2 +\frac{\Pi}{\Sigma}\sin^2{\vartheta}~\mathrm{d}\varphi^2 \right]\,.
\end{align}
Here the parameter $a$ is the spin of the central object and the ``stationary" metric functions $\{F, \Delta, \Sigma, \Pi\}$ can be related to the ``static" metric functions $\{f, g, R\}$ in \eqref{eq:Static_Spherical_Seed_Metric} as discussed below. This then leaves only the metric function $X = X(r, \vartheta)$ to be fixed by the field equations \citep{Azreg-Ainou2014b}.

While we focus in this paper on the AA metric, we note that different generalizations of the static metric \eqref{eq:Static_Spherical_Seed_Metric} are possible, in principle, and can be achieved by modifying the complex coordinate transformation involved in the \citet{Newman+1965a}-type solution-generating technique (see, e.g., \citealt{Azreg-Ainou2014a}). \RV{It is important to note, however, that the particular transformations used in \cite{Azreg-Ainou2014a, Azreg-Ainou2014b} send the Schwarzschild, the Reissner-Nordstr{\"o}m, and the GMGHS metrics, to their appropriate spinning generalizations, namely to the Kerr, the Kerr-Newman, and the Kerr-Sen metrics, respectively.} We direct the reader to \cite{Erbin2017} for a review of such solution-generating techniques, and to Sec.~7.1 of \cite{Wald1984} for a general discussion of stationary, axisymmetric spacetimes, and on the construction of BL-like coordinates. 

Expressions for the metric functions $\{F, \Delta, \Sigma, \Pi\}$ in the AA metric \eqref{eq:AA_Metric} are most conveniently written by first defining two
auxiliary functions that are specific combinations of the original static metric functions, $f(r)$, $g(r)$, $R(r)$,
\begin{align} \label{eq:AA_Auxiliary_Functions}
A(r) =\ R^2/\sqrt{g}\,,\  \
B(r) =\ (f/g)R^2\,, \nonumber
\end{align}
which can be written even more transparently as
\begin{equation} \label{eq:AA_Auxiliary_Functions_Gen_Form}
A = \frac{\hat{g}_{\vartheta\vartheta}}{\sqrt{-\hat{g}_{tt}\hat{g}_{rr}}}\,;\ \ B = \frac{\hat{g}_{\vartheta\vartheta}}{\hat{g}_{rr}} = \hat{g}^{rr}\hat{g}_{\vartheta\vartheta}\,.
\end{equation}
$A$ and $B$ are each functions of $r$ alone,  and in terms of them the functions
$\{F, \Delta, \Sigma, \Pi\}$ in the AA metric \eqref{eq:AA_Metric} take the remarkably simple form
\begin{align} \label{eq:AA_Metric_Functions}
F(r) =&\ (A - B)/2\,, \\
\Delta(r) =&\ B + a^2\,, \nonumber \\
\Sigma(r, \vartheta) =&\ A + a^2\cos^2{\vartheta}\,, \nonumber \\
\Pi(r, \vartheta) =&\ \left(A + a^2\right)^2 - \Delta a^2\sin^2{\vartheta}\,. \nonumber 
\end{align}
Note that $F$ and $\Delta$ depend only on $r$, while $\Sigma$ and $\Pi$ are functions of $r$ and $\vartheta$. The spinning AA metric $g_{\mu\nu}$ \eqref{eq:AA_Metric} reduces to the nonspinning metric $\hat{g}_{\mu\nu}$ \eqref{eq:Static_Spherical_Seed_Metric} in the limit of vanishing spin ($a=0$) for appropriate choices of $X$ (see \citealt{Azreg-Ainou2014a} for a discussion on conformal vs. normal fluids). 

\RV{The determinant $\det{[g_{\mu\nu}]}$ of the AA metric, and of its $t\varphi-$sector $\det{[g_{t\varphi}]} := g_{tt}g_{\varphi\varphi} - g_{t\varphi}^2$ are given in BL coordinates by
\begin{align}
\det{[g_{\mu\nu}]} =&\ -(X^4/\Sigma^2)\sin^2{\vartheta}\,,
\\
\det{[g_{t\varphi}]} =&\ -(X^2/\Sigma^2)\Delta\sin^2{\vartheta}\,. 
\end{align}
The AA metric can also be expressed in the closely-related quasi-isotropic (QI) coordinates (see, e.g., Sec. 2.3.2 of \citealt{Gourgoulhon2010}), $x^{\tilde{\mu}} = (t, \tilde{r}, \vartheta, \varphi)$, as
\begin{align} \label{eq:AA_Metric_QI}
\mathrm{d}s^2
=&\ \frac{X}{\Sigma}\left[
-\frac{X\Delta}{\Pi}\mathrm{d}t^2 + \frac{\Sigma}{\tilde{r}^2}\left(\mathrm{d}\tilde{r}^2 + \tilde{r}^2\mathrm{d}\vartheta^2\right) \right. \nonumber \\
&\ \left. + \frac{\Pi}{\Sigma}\sin^2{\vartheta}\left(\mathrm{d}\varphi-\Omega_{\mathrm{Z}}\mathrm{d}t\right)^2 \right]\,,
\end{align}
where we have introduced the angular velocity of the zero angular momentum observer (ZAMO), $\Omega_{\mathrm{Z}} = -g_{t\varphi}/g_{\varphi\varphi} = 2aF/\Pi$ (cf. eqs. 2.4, 2.5 of \citealt{Bardeen+1972}). The form above \eqref{eq:AA_Metric_QI} can be obtained from \eqref{eq:AA_Metric} by transforming the radial coordinate, $r \mapsto \tilde{r}$, via, 
$\tilde{r}(r) = \exp{[\int(\mathrm{d}r/\sqrt{\Delta})]}$. The $r\vartheta-$sector of the AA metric in these QI coordinates is clearly conformally flat, as desired.  The related cylindrical coordinates, $x^{\mu\prime} = (t, \rho, z, \varphi)$, with $\rho := \tilde{r}\sin{\vartheta}$ and $z := \tilde{r}\cos{\vartheta}$, are called the Lewis-Papapetrou coordinates \citep{Lewis1932, Papapetrou1966}.}

We briefly illustrate now how the remaining metric function $X$ can be fixed by considering, as an example, the case when the background matter is an anisotropic fluid in GR which flows around the spin axis. Since in its own rest-frame, $\{e^\mu_{(a)}\ (a=0-3)\}$, the fluid stress-energy-momentum tensor is diagonal, the Einstein equations imply that the Einstein tensor in this frame, $\mathscr{G}_{(a)(b)} = \mathscr{G}_{\mu\nu}e^\mu_{(a)}e^\nu_{(b)}$, must also be diagonal. However, it can be shown that two of the off-diagonal elements of the Einstein tensor, $\mathscr{G}_{(r)(\vartheta)}$ and $\mathscr{G}_{(t)(\varphi)}$, will in general be non-zero. Demanding that these two terms vanish yields one nonlinear partial differential equation (PDE) and one linear PDE, each of which involves the free metric function $X$, the auxiliary function $A(r)$, and the fluid angular velocity $\Omega$. \RV{These PDEs are given in eqs. 15 and 18 of \cite{Azreg-Ainou2014a}, respectively, where the fluid angular velocity was chosen to be $\Omega = a/(A+a^2)$ (see $e^\mu_t$ in eq. 16 there). It is interesting to note that this angular velocity does not correspond to the ZAMO angular velocity introduced above, $\Omega_{\mathrm{Z}} = 2aF/\Pi$. It matches, however, the angular velocity of the principal null congruences of the spacetime introduced below (see eq. \ref{eq:PNC_Vector_Fields_BL}).} For non-fluid matter models, additional equations of motion for the matter fields must be solved. For example, if the matter is a scalar field, the associated Klein-Gordon equations must additionally be solved. Similarly, if the matter is an electromagnetic field, the Maxwell equations have to be accounted for as well. We will not enter into a discussion of this topic but refer the reader to \cite{Erbin2017}.

For the purposes of the present paper, it suffices to note that the subclass of AA metrics with $X=\Sigma$ is particularly interesting, as it corresponds to spacetimes that are asymptotically flat (see Appendix \ref{sec:AppB_Asympt_Flat} for further details). Furthermore, it can be checked that this choice for the conformal factor $X$ always solves the $\mathscr{G}_{(r)(\vartheta)} = 0$ equation for axially-spinning matter. Finally, all of the BH solutions discussed in this paper have $X=\Sigma$, whether they are vacuum or contain scalar and/or electromagnetic and/or axion fields. Even though these spacetimes arise as ``solution metrics'' to a variety of field equations, $X=\Sigma=A(r) + a^2 \cos^2\vartheta$ consistently remains a valid choice. For this subclass of the AA metrics, the possibly divergent behavior of the Ricci $\mathscr{R}$ and the Kretschmann $\mathscr{K}$ scalars can be seen \RV{from the behavior of their denominators, which go roughly as} $\sim \Sigma^{-3}$ and as $\sim \Sigma^{-6}$ \RV{respectively}. Thus, when this metric describes a singular spacetime, the curvature singularity is located at \RV{$r=r_\star$ such that} $\Sigma(r_\star, \pi/2) = 0$, \RV{which corresponds to a ring singularity located in the equatorial plane. This last equation is equivalent to $R(r_\star) = 0$.}
    
When the general AA metric \eqref{eq:AA_Metric} describes a BH spacetime, as above, a horizon is present at every location $\RV{r=r_{\mathrm{H}} > r_\star}$ where $g^{rr} = \Delta/\Sigma = 0$. Equivalently, horizons are located at the real, positive ($\RV{R(r_{\mathrm{H}}) > 0}$) roots of $\Delta(r)$. \RV{Conversely, if no such roots exist, then the AA metric does not correspond to a BH spacetime (e.g., the $a>M$ Kerr metric). The event horizon in particular is located at the largest such root.} Since $\Delta$ is a function of $r$ alone, it is reassuring to find that the event horizon is indeed a round sphere in BL coordinates. \RV{Furthermore, the horizon angular velocity, $\Omega_{\mathrm{H}} := \Omega_{\mathrm{Z}}(r_{\mathrm{H}})$, is simply given as $\Omega_{\mathrm{H}} = 2aF(r_{\mathrm{H}})/(A(r_{\mathrm{H}})+a^2)^2 = a/(A(r_{\mathrm{H}})+a^2)$. Clearly, $\Omega_{\mathrm{H}}$ is independent of $\vartheta$, i.e., an AA BH rotates like a rigid body, as a consequence of the weak rigidity theorem (see, e.g., Sec. 8.4.4 of \citealt{Straumann2013}). These} desirable properties, which \RV{are} well-known for the Kerr and Kerr-Newman metrics, \RV{are} now seen to be true for the very wide class of AA stationary BH metrics.

\RV{It is interesting to note that the general AA metric \eqref{eq:AA_Metric} can also describe spacetimes which contain regions that admit closed timelike curves. In such regions, $g_{\varphi\varphi} < 0$  (see, e.g., Sec. V.B of \citealt{Johannsen2013a}), i.e., $\Pi < 0$ (e.g., the $a>M$ Kerr metric).}

\RV{We show in Appendix \ref{sec:AppC_Separability_AA_Metric} that} the null geodesic equations are Lioville-separable. \RV{Thus, the AA metric always admits a Carter constant for null geodesics.} Furthermore, we also identify a subclass of AA metrics that additionally admit separable timelike geodesic equations. \RV{These require the conformal factor to take the form $X(r,\vartheta) = X_r(r) + X_\vartheta(\vartheta)$. For this subclass, a Carter constant exists for all geodesics. Concomitantly, arbitrary geodesics of all AA metrics with $X=\Sigma$ possess Carter constants.}

The $rr$-component of the AA metric \eqref{eq:AA_Metric} in BL coordinates diverges at the horizon ($\Delta(r)=0$). The Ricci and Kretschmann scalars, however, reveal this to be merely a coordinate singularity, an artifact of the choice of coordinates. The singularity can be eliminated by a change of coordinates, and this is the topic we turn to next.

\section{Horizon-Penetrating Coordinates}
\label{sec:Sec4_siKS_Form_AA_Metric}

In this section, we convert the AA metric, which is written in BL coordinates in equation \ref{eq:AA_Metric}, to a horizon-penetrating form in which no coordinate singularity is present at the horizon. This is the form that is most useful for GRMHD and xGRMHD simulations of astrophysical accretion flows.

The ingoing ($-$) and outgoing ($+$) principal null congruences (PNCs) of the AA spacetime consist of null geodesics $x^\mu(\lambda)$ whose tangents $\ell_\pm^\mu = \dot{x}^\mu = \mathrm{d}x^\mu/\mathrm{d}\lambda$ satisfy $\ell_\pm^\vartheta = \dot{\vartheta} = 0$ and $\ddot{\vartheta} = 0$ (see, e.g., \citealt{Misner+1973}, \citealt{Hioki+2008} for further details). The $1-$form fields $(l_\pm)_\mu$ associated with the PNCs in an AA spacetime can be expressed elegantly as,%
\footnote{See Appendix \ref{sec:AppC_Separability_AA_Metric}. For completeness, the PNC vector fields are (cf. eq. 33.39 of \citealt{Misner+1973} for the Kerr metric),
\begin{equation} \label{eq:PNC_Vector_Fields_BL}
\ell_\pm^\mu = E\frac{\Sigma}{X}\left[\frac{A + a^2}{\Delta}, \pm 1, 0, \frac{a}{\Delta}\right]\,.
\end{equation}
\label{fn:FN2_Tangent_PNCs}}
\begin{equation} \label{eq:iPNC_1Form_BL}
\left(\ell_\pm\right)_\mu = E\left[-1, \pm \frac{\Sigma}{\Delta}, 0, a\sin^2{\vartheta}\right]\,,
\end{equation}
where $E$ is some constant.

We now define the spherical-polar ingoing Kerr-Schild (siKS) coordinates, $x^{\bar{\mu}} = (\tau, r, \vartheta, \phi)$, as those in which the ingoing PNC takes the form, 
\begin{equation} \label{eq:iPNC_1Form_siKS}
\left(\ell_-\right)_{\bar{\mu}} = E\left[-1, -1, 0, a\sin^2{\vartheta}\right]\,.
\end{equation}
To distinguish between the BL ($x^\mu$) and the siKS ($x^{\bar{\mu}}$) coordinate systems, we will use a bar for the indices of the latter. We are interested in the ingoing PNC since our coordinate system should penetrate the future horizon $\mathscr{H}^+$ when one exists, and cover patches I and II of the Kruskal \citep{Kruskal1960} or Penrose-Carter \citep{Penrose1963, Carter1966} diagrams. The Jacobian $\Lambda^{\bar{\mu}}_{\ \mu} = \partial_\mu x^{\bar{\mu}}$ for the coordinate transformation from BL coordinates to siKS coordinates (i.e., $\mathrm{d}x^{\bar{\mu}} = \Lambda^{\bar{\mu}}_{\ \mu}\mathrm{d}x^\mu$) can be inferred from equations \ref{eq:iPNC_1Form_BL} and \ref{eq:iPNC_1Form_siKS} to be,
\begin{equation} \label{eq:Jacobian_BL_siKS}
\Lambda^{\bar{\mu}}_{\ \mu} = 
\begin{bmatrix}
1 & 2 F/\Delta & 0 & 0 \\
0 & 1 & 0 & 0 \\
0 & 0 & 1 & 0 \\
0 & a/\Delta & 0 & 1
\end{bmatrix}\,.
\end{equation}
The time coordinate $\tau$ then is clearly a ``tortoise'' time coordinate (see, e.g., \citealt{Blau2023}),
\begin{equation}
\tau = t + \int (2F/\Delta)~\mathrm{d}r\,,
\end{equation}
and we recognize immediately that, in the limit of vanishing spin ($a \rightarrow 0$), the siKS coordinates are analogous to the coordinates presented in \cite{Eddington1924} and \cite{Finkelstein1958}.

Using the above coordinate transformation, we can write the AA metric in siKS coordinates as,
\begin{equation} \label{eq:AA_Metric_siKS_Coordinates}
g_{\bar{\mu}\bar{\nu}} = \frac{X}{\Sigma}
\begin{bmatrix}
-\left(1 - \frac{2 F}{\Sigma}\right) & \frac{2 F}{\Sigma} & 0 & -\left(\frac{2 F}{\Sigma}\right)a\sin^2{\vartheta} \\ 
* & \left(1+\frac{2 F}{\Sigma}\right) & 0 & -\left(1+\frac{2 F}{\Sigma}\right)a\sin^2{\vartheta} \\
* & * & \Sigma & 0 \\
* & * & * & \left(\frac{\Pi}{\Sigma}\right)\sin^2\vartheta
\end{bmatrix},
\end{equation}
where the asterisks denote components that are fixed by symmetry, $g_{\bar{\mu}\bar{\nu}} = g_{\bar{\nu}\bar{\mu}}$. Notice that the metric components now diverge only at the spacetime/curvature singularity, present at $\Sigma = 0$. \RV{The inverse AA metric in siKS coordinates is
\begin{equation} \label{eq:AA_Inverse_Metric_siKS_Coordinates}
g^{\bar{\mu}\bar{\nu}} = \frac{\Sigma}{X}
\begin{bmatrix}
-\left(1 + \frac{2 F}{\Sigma}\right) & \frac{2 F}{\Sigma} & 0 & 0 \\ 
* & \frac{\Delta}{\Sigma} & 0 & \frac{a}{\Sigma} \\
* & * & \frac{1}{\Sigma} & 0 \\
* & * & * & \frac{1}{\Sigma\sin^2\vartheta}
\end{bmatrix}\,.
\end{equation}
For completeness, we cast the general AA metric into its ``classic'' Kerr-Schild form in eq. \ref{eq:AA_Metric_Kerr_Schild_Form} of  Appendix \ref{sec:AppB_Asympt_Flat}.}

\RV{In siKS coordinates,} the event horizon is still located at the outermost root of $g^{\bar{1}\bar{1}}=g^{rr}=0$, i.e., $\Delta(r)=0$ \eqref{eq:AA_Inverse_Metric_siKS_Coordinates}, but as we can see from eq. \ref{eq:AA_Metric_siKS_Coordinates} the metric components $g_{\bar{\mu}\bar{\nu}}$ no longer diverge at $\Delta=0$. Thus, the coordinate singularity at the horizon has been removed by changing from BL to siKS coordinates, showing that the latter are horizon-penetrating coordinates. Furthermore, a comparison of the Kerr metric in particular, in these coordinates, with that reported in \cite{Font+1998}, \RV{with $k=1$ there}, and in \cite{McKinney+2004}, shows that these are the horizon-penetrating coordinates canonically used in the context of GRMHD simulations. 

Note that the original set of Kerr-Schild coordinates used to describe the Kerr spacetime \citep{Kerr1963, Kerr+2009} are adapted to the outgoing PNC in which $l_+^{\breve{\mu}} \propto \delta^{\breve{\mu}}_r$ (see, e.g., \citealt{Wiltshire+2009, Azreg-Ainou2014b}). The corresponding time coordinate $u$ is the retarded time as well as a null affine parameter that parametrizes the outgoing PNC on future null infinity $\mathscr{I}^+$. Equivalently, the ingoing version of the original set of Kerr-Schild coordinates would be adapted to the ingoing PNC in which $l_-^{\breve{\mu}} \propto \delta^{\breve{\mu}}_r$ with the corresponding time coordinate $v$ being the advanced time as well as a null affine parameter that parametrizes the ingoing PNC on past null infinity $\mathscr{I}^-$. Clearly, the ``original'' ingoing Kerr-Schild system is quite different from the spherical-ingoing Kerr-Schild coordinates which we use \RV{in this section} and which are matched to current GRMHD codes. From eq. \ref{eq:iPNC_1Form_siKS}, that the tangent to the ingoing PNC in our coordinates is given as
\begin{equation}
\ell_-^{\bar{\mu}} = E\frac{\Sigma}{X}\left[1, -1, 0, 0\right]\,.
\end{equation}

\section{Horizon-Penetrating $3+1$ Form}
\label{sec:Sec4.1_3+1_Form}
We can cast the AA metric when written in horizon-penetrating siKS coordinates into its associated $3+1$ form \citep{Arnowitt+2008} by introducing a foliation into spacelike hypersurfaces $\Sigma_\tau$ that are iso-surfaces of the scalar time function $\tau$ (see, e.g., \citealt{Poisson2004}, \citealt{Gourgoulhon2007}, \citealt{Alcubierre2008}, \citealt{Rezzolla+2013}). The unit timelike normal to these hypersurfaces is
\begin{equation} \label{eq:Hypersurface_Normal}
n_{\bar{\mu}} := -\alpha\nabla_{\bar{\mu}} \tau \RV{= -\alpha\delta_{\bar{\mu}}^\tau}\,,
\end{equation}
where $\alpha := \sqrt{-1/g^{\tau\tau}}$ is called the lapse function, \RV{and from eq. \ref{eq:AA_Inverse_Metric_siKS_Coordinates} is given by
\begin{equation}
\alpha = [X/(\Sigma+2F)]^{1/2}\,.
\label{eq:lapse}
\end{equation}
The four-vector that is dual to the normal, i.e., $n^{\bar{\mu}} = g^{\bar{\mu}\bar{\nu}}n_{\bar{\nu}} = -\alpha g^{\bar{\mu}\tau} = [1/\alpha, -\alpha(2F/X), 0, 0]$, corresponds to the four-velocity of the local timelike Eulerian observer. The four-velocity of this observer in BL coordinates can be obtained as $n^\mu = \Lambda^\mu_{\ \bar{\mu}}n^{\bar{\mu}}$, where $\Lambda^\mu_{\ \bar{\mu}} = \left(\Lambda^{\bar{\mu}}_{\ \mu}\right)^{-1}$. It is straightforward to then check that the angular velocity of the Eulerian observer in BL coordinates $\Omega_{\mathrm{E}} := n^\varphi/n^t$ matches the angular velocity of the zero angular momentum observer introduced above $\Omega_{\mathrm{Z}} = -g_{t\varphi}/g_{\varphi\varphi} = 2aF/\Pi$ exactly. However, the Eulerian observer has nonzero BL radial velocity, i.e., $n^r = -2\alpha F/X (= n^{\bar{r}})$.}

With eq. \ref{eq:Hypersurface_Normal}, the induced (Riemannian) metric $\gamma$ on the spacelike hypersurfaces can now be introduced as
\begin{equation}
\gamma_{\bar{\mu}\bar{\nu}} := g_{\bar{\mu}\bar{\nu}} + n_{\bar{\mu}}n_{\bar{\nu}}\,.
\label{eq:induced_metric}
\end{equation}
\RV{The projection tensor, which is used to obtain the spatial components of any four-vector or tensor, is simply $\gamma^{\bar{\mu}}_{\ \bar{\nu}} = g^{\bar{\mu}\bar{\alpha}}\gamma_{\bar{\alpha}\bar{\nu}} = \delta^{\bar{\mu}}_{\ \bar{\nu}} + n^{\bar{\mu}}n_{\bar{\nu}}$ since $\gamma^{\bar{\mu}}_{\ \bar{\nu}}n^{\bar{\nu}} = 0$.}

\RV{Although the 1-form $n_{\bar{\mu}}\mathrm{d}x^{\bar{\mu}}$ is collinear with $\mathrm{d}\tau$ (see eq. \ref{eq:Hypersurface_Normal}), the dual four-vector $n^{\bar{\mu}}\partial_{\bar{\mu}}$ is not, in general, collinear with $\partial_\tau$ (since $-\alpha g^{\bar{\mu}\tau} \cancel{\propto}\ \delta^{\bar{\mu}}_\tau$). Indeed, this noncollinearity is typically captured via a four-vector $\beta^{\bar{\mu}}$, which is defined such that $n_{\bar{\mu}}\beta^{\bar{\mu}} = 0$:
\begin{equation}
\beta^{\bar{\mu}} := (\partial_\tau)^{\bar{\mu}} - \alpha n^{\bar{\mu}} = \delta^{\bar{\mu}}_\tau - \alpha n^{\bar{\mu}}\,.
\end{equation}
Since by construction $\beta^\tau = 0$,} it is convenient to work instead with the spacelike three-vector (see, e.g., \citealt{McKinney+2004}, \citealt{Porth+2017})
\begin{equation} 
\beta^i := \alpha^2g^{i\tau} = -g^{i\tau}/g^{\tau\tau} \RV{= [2F/(2F+\Sigma)]\delta^{i}_{r}}\,,
\label{eq:shift}
\end{equation}
where the index $i$ takes values $i=1, 2, 3$. \RV{This three-vector measures the shift of the isolines of the three spatial coordinates with respect to the normal to the hypersurfaces and is therefore called the shift vector (see, e.g., Sec. 1.3 of \citealt{Gourgoulhon2010}).}

The desired $3+1$ form of the AA metric can now be written, using the lapse function $\alpha$ \RV{(eq. \ref{eq:lapse})}, the shift vector $\beta^i$ \RV{(eq. \ref{eq:shift})}, and the components $\gamma_{ij}$ \RV{(eq. \ref{eq:induced_metric})} of the induced metric on the hypersurfaces $\Sigma_\tau$, as
\begin{equation}
\mathrm{d}s^2 = -\alpha^2\mathrm{d}\tau^2 + \gamma_{ij}(\mathrm{d}x^i + \beta^i\mathrm{d}\tau)(\mathrm{d}x^j + \beta^j\mathrm{d}\tau)\,.
\end{equation}
\RV{For completeness, we demonstrate how the temporal $u^\tau$ and the spatial $u^i$ components of a vector $u^{\bar{\mu}}=(u^\tau, u^i)$ can be obtained using the unit normal and the projection tensor respectively. First, with $\Gamma = -n_{\bar{\mu}}u^{\bar{\mu}}$, we have $u^\tau = \Gamma/\alpha$. Second, introducing the purely spatial vectors ($v^0 = 0$) $v^i := (\gamma^i_{\ \bar{\mu}}u^{\bar{\mu}})/\Gamma$, we obtain $u^i = \Gamma(v^i - \beta^i/\alpha)$. One also finds that $\Gamma = 1/\sqrt{1-v^2}$, where $v^2 = v_iv^i$. When $u$ corresponds to the four-velocity of a fluid element, $\Gamma$ has the significance of being the fluid Lorentz factor with respect to the Eulerian observer. An excellent summary of the 3+1 treatment relevant for GRMHD can be found in \cite{Gammie+2003}.}

\section{Summary and Conclusions}
\label{sec:Sec5_Conclusions}

We have presented in equation \eqref{eq:AA_Metric_siKS_Coordinates}, section \ref{sec:Sec4_siKS_Form_AA_Metric}, a horizon-penetrating Kerr-Schild form of the rather general stationary and axisymmetric Azreg-A{\"i}nou (AA) metric. This form of the metric can be used as an input when performing general relativistic (GR) as well as non-GR (xGR) magnetohydrodynamics simulations of accretion flows onto compact objects in arbitrary metric theories of gravity. The AA metric can be used to generate a number of popular metrics that describe the stationary spacetimes in GR corresponding to electro-vacuum black holes \citep{Kerr1963, Newman+1965b}, regular black holes such as the Kerr-Hayward models \citep{Hayward2006, Bambi+2013, Zhou+2023a, Zhou+2023b}, naked singularities such as the Kerr-Janis-Newman-Winicour spacetime \citep{Solanki+2022}, etc. It can also be used to describe black hole solutions arising in alternative theories of gravity (e.g., \citealt{Sen1992}). 

\begin{table*}[ht!]
\begin{center}
\caption{
\textit{Metric functions for the stationary and axisymmetric generalizations of the static and spherically-symmetric spacetimes given in Table \ref{table:Known_Static_Solutions}.} 
The spinning generalizations of a nonspinning ``seed'' metric \eqref{eq:Static_Spherical_Seed_Metric} is given in Boyer-Lindquist coordinates in eq. \ref{eq:AA_Metric} and in horizon-penetrating spherical ingoing Kerr-Schild coordinates in eq. \ref{eq:AA_siKS_C}. These involve five ``stationary metric functions'' $\{F, \Delta, \Sigma, \Pi, X\}$, the first four of which are related to $f$ and $R$ via eq. \ref{eq:AA_Staionary_Metric_Functions_fgh}. We show below the spinning generalizations of the specific static solutions listed in Table \ref{table:Known_Static_Solutions}. Since we used coordinates for the static metrics in which $g(r)=1$, $g$ does not appear in the expressions below. For brevity we only show the four metric functions $2F$, $\Delta$, $\Sigma=X$ below. The fifth metric function is given by $\Pi = (R^2 + a^2)^2 -\Delta a^2 \sin^2{\vartheta}$. All the spacetime models listed here are asymptotically flat, and $a$ denotes the usual spin parameter. A variety of physical scenarios are captured by the exemplary set listed here: the vacuum Kerr BH, the electromagnetically charged Kerr-Newman BH, the (modified) Kerr-Hayward spinning BH, the spinning JNW naked singularity, and the spinning JMN-1 naked singularity are solutions of general relativity, whereas the Kerr-Sen spacetime describes electromagnetically-charged BHs of string theory which are also charged under a dilaton field as well as an axion field.}
\label{table:siKS_Known_Solutions}
\centering
\renewcommand{\arraystretch}{2.5}
\begin{tabular}[t]{|l|c|c|c|}
\hline
\textbf{Spacetime} 
& $2F = (1-f)R^2$
& $\Delta = f R^2 + a^2$ 
& $\Sigma = X = R^2 + a^2\cos^2{\vartheta}$
\\
\hline
Kerr 
& $2Mr$ 
& $r^2-2Mr+a^2$ 
& $r^2+a^2\cos^2{\vartheta}$ 
\\ 
Kerr-Newman 
& $2Mr - Q^2$ 
& $r^2 - 2Mr+ Q^2+a^2$ 
& $r^2+a^2\cos^2\vartheta$ 
\\
Kerr-Hayward 
& $\frac{2Mr^5}{r^4 + 2L^4}$ 
& $r^2 - \frac{2Mr^5}{r^4 + 2L^4} + a^2$ 
& $r^2+a^2\cos^2{\vartheta}$ 
\\
Kerr-Sen 
& $2Mr-2Q^2$ 
& $r^2-\left(2M+\frac{Q^2}{M}\right)r+2Q^2+a^2$ 
& $r\left(r-\frac{Q^2}{M}\right) + a^2\cos^2{\vartheta}$ 
\\ 
Spinning JNW 
& $r^2\left[\left(1-\frac{r_\star}{r}\right)^{\hat{\nu}} - \left(1-\frac{r_\star}{r}\right)\right]$ 
& $r^2 - \frac{2Mr}{1-\hat{\nu}} + a^2$ 
& $r^2\left(1-\frac{r_\star}{r}\right)^{\hat{\nu}} + a^2\cos^2{\vartheta}$ 
\\
Spinning JMN-1 
& $\frac{r^2}{(1-\sigma)^2}\left[\left(\frac{r}{r_{\mathrm{b}}}\right)^{-2\sigma} - \left(1-\frac{2M}{r_{\mathrm{b}}}\right)\right]$
& $\left(1-\frac{2M}{R_{\mathrm{b}}}\right)\frac{r^2}{(1-\sigma)^2} + a^2$
& $R_{\mathrm{b}}^2\left(\frac{r}{r_{\mathrm{b}}}\right)^{2-2\sigma}+a^2\cos^2{\vartheta}$ 
\\
\hline
\end{tabular}
\end{center}
\end{table*}

To illustrate the ease of generating the Kerr-Schild forms of new stationary metrics, we now summarize the steps involved. We start with an arbitrary static and spherically-symmetric ``seed'' metric, e.g., the ones listed in Table~\ref{table:Known_Static_Solutions}, in standard spherical-polar coordinates, $x^\mu = (t, r, \vartheta, \varphi)$,
\begin{equation} \label{eq:Static_Spherical_Seed_Metric_fgR}
\mathrm{d}s^2 = -f(r)\mathrm{d}t^2 + \frac{g(r)}{f(r)}\mathrm{d}r^2 + R^2(r)\mathrm{d}\Omega_2^2\,.
\end{equation}
As we noted in section \ref{sec:Sec2_Static_Metric}, it is always possible to find a radial coordinate $r$ in which $g(r)=1$. Since this reduces the number of free functions, we used such coordinates for the static models listed in Table \ref{table:Known_Static_Solutions}. However, the required transformation may not always lead to simple analytical forms for $f(r)$ and $R(r)$. When it does not, it is preferable to use the more general form \eqref{eq:Static_Spherical_Seed_Metric_fgR}.

The AA stationary generalization of the metric \eqref{eq:Static_Spherical_Seed_Metric_fgR} in spherical-polar ingoing Kerr-Schild coordinates, $x^{\bar{\mu}} = (\tau, r, \vartheta, \phi)$, takes the form,
\begin{align} \label{eq:AA_siKS_C}
\mathrm{d}s^2 =&\ \frac{X}{\Sigma}\left[-\left(1-\frac{2F}{\Sigma}\right)\mathrm{d}\tau^2 +  \left(1+\frac{2F}{\Sigma}\right)\mathrm{d}r^2 \right. \nonumber \\
&\ \left. + \Sigma~\mathrm{d}\vartheta^2  + \frac{\Pi}{\Sigma}\sin^2{\vartheta}~\mathrm{d}\phi^2 
- 2\frac{2F}{\Sigma}a\sin^2{\vartheta}~\mathrm{d}\tau\mathrm{d}\phi \right. \nonumber \\
&\ \left. + 2\frac{2F}{\Sigma}\mathrm{d}\tau\mathrm{d}r - 2\left(1+\frac{2F}{\Sigma}\right)a\sin^2{\vartheta}~\mathrm{d}r\mathrm{d}\phi\right]\,,
\end{align}
where the ``stationary metric functions'' $\{F, \Delta, \Sigma, \Pi\}$ are fixed by the ``static metric functions'' $\{f, g, R\}$ in the seed metric \eqref{eq:Static_Spherical_Seed_Metric_fgR} as,
\begin{align} \label{eq:AA_Staionary_Metric_Functions_fgh}
2F(r) =&\ R^2/\sqrt{g}-(f/g)R^2\,, \\ 
\Delta(r) =&\ (f/g)R^2 + a^2\,, \nonumber \\
\Sigma(r, \vartheta) =&\ R^2/\sqrt{g} + a^2\cos^2{\vartheta}\,,\nonumber \\ 
\Pi(r, \vartheta) =&\ (R^2/\sqrt{g}+a^2)^2 - \Delta a^2\sin^2{\vartheta}\,. \nonumber 
\end{align}
\RV{The parameter $a$ describes the spin of the spacetime.} The only remaining freedom is the stationary metric function $X(r,\vartheta)$.

If the background matter in the stationary spacetime is a fluid flowing around the spin axis ($\vartheta=0$) in general relativity (GR), the unknown function $X=X(r,\vartheta)$ is fixed by solving two Einstein equations, $\mathscr{G}_{(r)(\vartheta)}=0$ and $\mathscr{G}_{(t)(\varphi)}=0$, in the fluid rest frame (the energy-momentum-stress components $T_{(r)(\vartheta)}$ and $T_{(t)(\varphi)}$ of the fluid vanish in this frame). We will not enter into detailed implications of these equations here \RV{but note that $X=\Sigma$ is always a solution of the first equation, $\mathscr{G}_{(r)(\vartheta)}=0$. The choice $X=\Sigma$ is additionally appealing since spacetimes with this property} are asymptotically-flat (provided of course that the original seed static metric is also asymptotically flat). Verifying that $X=\Sigma$ also solves the second Einstein equation, $\mathscr{G}_{(t)(\varphi)}=0$, in the fluid frame, and that the resulting energy-momentum-stress tensor is physically valid, i.e., satisfies various energy conditions, is an involved calculation and is beyond the scope of this paper. However, if one merely wants a novel stationary metric in which to carry out GRMHD simulations, and if one is not too concerned about necessarily satisfying energy conditions, then the metric \eqref{eq:AA_siKS_C} with $X=\Sigma$, and a suitable choice of the seed functions $f(r)$, $g(r)$, $R(r)$, would be a reasonable choice.

We should note that, for non-fluid models, additional equations of motion for the fields (e.g., the Klein-Gordon equations or the Maxwell equations) must be solved. The status and successes of such \textit{metric-}solution-generating techniques in being able to yield solutions also to the nongravitational field equations has been discussed in \cite{Erbin2017}.

Table \ref{table:siKS_Known_Solutions} presents a compilation of the metric functions of the stationary AA spacetimes corresponding to the six static ``seed" models listed in Table~\ref{table:Known_Static_Solutions}. These are all models for which the choice $X=\Sigma$ is valid. The stationary \cite{Kerr1963} metric arises naturally as the AA generalization of the static Schwarzschild metric. It describes vacuum, spinning black holes (BHs) in GR and has been used extensively in GRMHD simulations. Similarly, the AA transformation automatically generates the Kerr-Newman metric \citep{Newman+1965b}, which describes electromagnetically-charged, spinning BHs in GR, as the spinning generalization of the non-spinning Reissner-Nordstrom BH. The Kerr-Hayward model proposed here (the third model in Table~\ref{table:siKS_Known_Solutions}) is expected to describe spinning regular (i.e., no singularities) BHs in GR (see, e.g., \citealt{Bambi+2013, Zhou+2023b}). The spinning equivalent of the GMGHS BH solution in string theory is given by the Kerr-Sen metric \citep{Sen1992}. The form of the Kerr-Sen metric shown in Table~\ref{table:siKS_Known_Solutions} can be put into the form presented in eq. 20 of \cite{Xavier+2020} by replacing $r$ with $r+Q^2/M$. It is worth noting that the spacetime of Kerr-Sen BHs contains not just electromagnetic fields but also a scalar field (dilaton) as well as an axion field. All of the BH solutions in Table~\ref{table:siKS_Known_Solutions} reduce to the Kerr BH in appropriate limits ($Q\to0$, $L\to0$). 

The spinning Janis-Newman-Winicour (JNW) spacetime (the fifth model in Table~\ref{table:siKS_Known_Solutions}) has been proposed in \cite{Solanki+2022} and is expected to describe a spinning naked singularity spacetime containing scalar field matter. Finally, the spinning JMN-1 metric proposed here is expected to describe a spinning naked singularity spacetime containing an anisotropic fluid in GR. Careful explorations of these spacetimes, especially with a focus on the physicality of the underlying matter (e.g., how well energy conditions are satisfied), \RV{is left for future work.}

\cite{Vagnozzi+2022} give an immense compilation of spacetime metrics including those corresponding to static and spherically-symmetric spacetimes. These metric functions describe alternative nonspinning black holes which are of considerable interest and can be studied using simulations. More importantly, each static model could potentially be used as a seed metric to obtain the horizon-penetrating form of the stationary metric that it may belong to via equations \ref{eq:AA_siKS_C} and \ref{eq:AA_Staionary_Metric_Functions_fgh}. The only open issue, which needs to be checked in each case, is whether the choice $X=\Sigma$ is physically valid.

We conclude on an optimistic note, envisioning the widespread adoption of GRMHD simulations, facilitated by our easy-to-use metric formulation. A Kerr simulation library has recently been used successfully to infer the properties of plasma present in the close vicinity of Sgr A$^*$, as well as the spacetime geometry of this supermassive BH \citep{EHTC+2022e, EHTC+2022f}. While it might be a challenging endeavor in the near future to construct similar libraries of fully 3D simulations in the other stationary spacetimes described in this paper (because of the many additional spacetime parameters in these models such as $Q$, $L$, $\hat{\nu}$, $\sigma$), it should at least be possible to build extensive 2D GRMHD simulation libraries for these spacetimes. This will enable valuable exploration of magnetized relativistic gas dynamics in diverse spacetimes.

\begin{acknowledgments}
We thank the referee for helpful suggestions and Saurabh for pointing out a typo. PK, RN, and KC acknowledge support in part from grants from the Gordon and Betty Moore Foundation and the John Templeton Foundation to the Black Hole Initiative at Harvard University, and from NSF award OISE-1743747. YM is supported by the National Natural Science Foundation of China (Grant No. 12273022) and Shanghai target program of basic research for international scientists (Grant No. 22JC1410600).
\end{acknowledgments}


\appendix

\section{Alternative Form for Arbitrary Static and Spherically-Symmetric Metrics}
\label{sec:AppA_g1_Form}

Given an arbitrary spherically-symmetric and static metric $\hat{g}_{\mu\nu}$ in arbitrary spherical-polar coordinates of the form \eqref{eq:Static_Spherical_Seed_Metric},
\begin{equation} 
\mathrm{d}s^2 =\ -f(r)\mathrm{d}t^2 + \frac{g(r)}{f(r)}\mathrm{d}r^2 + R^2(r)\mathrm{d}\Omega_2^2\,,
\end{equation}
we can put it into the form where $g(r)=1$ as,
\begin{align} \label{eq:g1_Form_rho}
\mathrm{d}s^2 &=\ -f(r)\mathrm{d}t^2 + \frac{g(r)}{f(r)}\left(\partial_\rho r\right)^2\mathrm{d}\rho^2 + R^2(r)\mathrm{d}\Omega_2^2\,, \nonumber \\
&=\ -f(\rho)\mathrm{d}t^2 + \frac{\mathrm{d}\rho^2}{f(\rho)} + R^2(\rho)\mathrm{d}\Omega_2^2\,,
\end{align}
where in writing the last equality we have imposed the condition that $r(\rho)$ satisfy the ordinary differential equation,
\begin{equation} \label{eq:g1_Coordinate_Transformation}
\sqrt{g(r)}~\partial_\rho r = 1\,,
\end{equation}
and $f(\rho) = f(r(\rho))$. Note that the above equation can be rewritten as
\begin{equation} \label{eq:g1_Coordinate_Transformation_Gen}
\sqrt{-\hat{g}_{tt}(r)\hat{g}_{rr}(r)}~\partial_\rho r = 1\,.
\end{equation}
Relabeling $\rho$ by $r$ in eq. \ref{eq:g1_Form_rho}, we have,
\begin{equation} \label{eq:g1_Form}
\mathrm{d}s^2 =\ -f(r)\mathrm{d}t^2 + \frac{\mathrm{d}r^2}{f(r)} + R^2(r)\mathrm{d}\Omega_2^2\,.
\end{equation}

We now demonstrate how we can straightforwardly obtain the JMN-1 naked singularity metric in the form given above \eqref{eq:g1_Form}, where $-\hat{g}_{tt}\hat{g}_{rr} = 1$. Its original form, as given in eq. 29 of \cite{Joshi+2011}, uses areal-radial coordinates, $x^\mu=(t, R, \vartheta, \varphi)$, in which $-\hat{g}_{tt}\hat{g}_{RR} \neq 1$. From the original form,
\begin{equation} \label{eq:JMN1_Original}
\mathrm{d}s^2 =\ -\left(1 - M_0\right)\left(\frac{R}{R_{\mathrm{b}}}\right)^{\frac{M_0}{1-M_0}}\mathrm{d}t^2 + \frac{\mathrm{d}R^2}{\left(1 - M_0\right)} + R^2~\mathrm{d}\Omega_2^2\,, \nonumber
\end{equation}
where the compactness parameter $M_0$ is given in terms of the ADM mass $M$ and the physical boundary or matching areal-radius $R_{\mathrm{b}}$ as $M_0 = 2M/R_{\mathrm{b}}$, we can write the desired equation \eqref{eq:g1_Coordinate_Transformation_Gen} to put it in the ``$g(r)=1$'' coordinates \eqref{eq:g1_Form}  used in Table \ref{table:Known_Static_Solutions} as,
\begin{align}
\left(\frac{R}{R_{\mathrm{b}}}\right)^\alpha\mathrm{d}R = \mathrm{d}r\,,
\end{align}
where $\alpha=M_0/(2-2M_0)$, \RV{not to be confused with the lapse function}, which yields a solution,
\begin{equation} \label{eq:rR_JMN1}
\frac{R_{\mathrm{b}}}{\alpha+1}\left(\frac{R}{R_{\mathrm{b}}}\right)^{\alpha+1} = r + k\,.
\end{equation}
The integration constant $k$ above can be set to zero so that at $R = 0$ we also have $r = 0$. The above equation is easily invertible as,
\begin{equation} \label{eq:Rr_JMN1}
R(r) = R_{\mathrm{b}}\left[\left(\frac{\alpha+1}{R_{\mathrm{b}}}\right)r\right]^{\frac{1}{\alpha+1}}\,.
\end{equation}
Now, with $M_0 = 2\alpha/(1+2\alpha)$, and 
\begin{equation}
\partial_r R(r) = \left[\left(\frac{\alpha+1}{R_{\mathrm{b}}}\right)r\right]^{\frac{-\alpha}{\alpha+1}}\,,
\end{equation}
we can rewrite the metric given in eq. \ref{eq:JMN1_Original} as,
\begin{align} \label{eq:JMN1_alpha}
\mathrm{d}s^2 &=\ -\left(\frac{1}{1+2\alpha}\right)\left[\left(\frac{\alpha+1}{R_{\mathrm{b}}}\right)r\right]^{\frac{2\alpha}{\alpha+1}}\mathrm{d}t^2 \\
&\ + \left(1+2\alpha\right)\left[\left(\frac{\alpha+1}{R_{\mathrm{b}}}\right)r\right]^{\frac{-2\alpha}{\alpha+1}}\mathrm{d}r^2 \nonumber \\
&\ + R_{\mathrm{b}}^2\left[\left(\frac{\alpha+1}{R_{\mathrm{b}}}\right)r\right]^{\frac{2}{\alpha+1}}~\mathrm{d}\Omega_2^2\,. \nonumber
\end{align}
Introducing $\sigma$ and $r_{\mathrm{b}}$,
\begin{align}
\sigma &=\ \frac{\alpha}{\alpha+1} = \frac{M}{R_{\mathrm{b}} - M}\,, \\
r_{\mathrm{b}} &=\ (1-\sigma)R_{\mathrm{b}} = \left(\frac{R_{\mathrm{b}}-2M}{R_{\mathrm{b}}-M}\right)R_{\mathrm{b}}\,, \nonumber
\end{align}
so that, $\alpha = \sigma/(1-\sigma)$, we can simplify eq. \ref{eq:JMN1_alpha} and write instead,
\begin{align} \label{eq:JMN1_sigma}
\mathrm{d}s^2 &=\ -\left(1-\frac{2M}{R_{\mathrm{b}}}\right)\left(\frac{r}{r_{\mathrm{b}}}\right)^{2\sigma}\mathrm{d}t^2 \\
&\ + \left(1-\frac{2M}{R_{\mathrm{b}}}\right)^{-1}\left(\frac{r}{r_{\mathrm{b}}}\right)^{-2\sigma}\mathrm{d}r^2 \nonumber \\
&\ + R_{\mathrm{b}}^2\left(\frac{r}{r_{\mathrm{b}}}\right)^{2-2\sigma}\mathrm{d}\Omega_2^2\,. \nonumber
\end{align}
This is the form of the JMN-1 metric reported in Table \ref{table:Known_Static_Solutions}. We note that while inverting $r(R)$, given in eq. \ref{eq:rR_JMN1}, to $R(r)$, given in eq. \ref{eq:Rr_JMN1}, was trivial in this case, this step might not be analytically feasible in general (assuming eq. \ref{eq:g1_Coordinate_Transformation_Gen} admits a closed-form solution in the first place). Therefore, it is convenient to use three metric functions, $f, g, R$, in general, as in eq. \ref{eq:Static_Spherical_Seed_Metric}.

\section{A Subclass of Asymptotically-Flat Stationary Metrics}
\label{sec:AppB_Asympt_Flat}

Since we are focused here on asymptotically-flat spacetimes (see, e.g., \citealt{Adamo+2009}), let us note first that for the seed metric \eqref{eq:Static_Spherical_Seed_Metric} to be asymptotically-flat, its metric functions must become $R(r) = r$, $f(r) = 1$, and $g(r)=1$ in the limit $r\rightarrow\infty$. Then, if we define $\tilde{\eta}_{\bar{\mu}\bar{\nu}}$ as,
\begin{equation} \label{eq:Asympt_Minkowski_siKS}
\tilde{\eta}_{\bar{\mu}\bar{\nu}} = 
\begin{bmatrix}
-1 & 0 & 0 & 0 \\
0 & 1 & 0 & -a\sin^2{\vartheta} \\
0 & 0 & A + a^2\cos^2{\vartheta} & 0 \\
0 & -a\sin^2{\vartheta} & 0 & (A+a^2)\sin^2{\vartheta}
\end{bmatrix}\,,
\end{equation}
we will find that as $r\rightarrow\infty$, the Riemann tensor associated with $\tilde{\eta}$ vanishes identically for such asymptotically-flat seeds. Thus, asymptotically, $\tilde{\eta}$ is the flat Minkowski metric $\eta$, in disguise, i.e., $\lim_{r\rightarrow\infty}\tilde{\eta} = \eta$. We stress that the Riemann tensor does \textit{not} vanish in general at finite coordinate radii $r$ for the metric tensor $\tilde{\eta}$.

Now, the AA metric in the siKS coordinates \eqref{eq:AA_Metric_siKS_Coordinates} can be expressed in terms of $\tilde{\eta}$ and the tangent to the ingoing PNC $\ell_{-}$ \eqref{eq:iPNC_1Form_siKS} everywhere as,
\begin{equation} \label{eq:AA_Metric_Kerr_Schild_Form}
g_{\bar{\mu}\bar{\nu}} = \frac{X}{\Sigma}\left[\tilde{\eta}_{\bar{\mu}\bar{\nu}} + \frac{2F}{\Sigma}(\ell_-)_{\bar{\mu}}(\ell_-)_{\bar{\nu}}\right]\,.
\end{equation}
We note further that $\ell_{-}$ is null with respect to both $g$ and $\tilde{\eta}$. 
We refer to the form of the metric above as the generalized (spherical ingoing) Kerr-Schild form of the AA metric (see Sec. 32.5 of \citealt{Stephani+2009} for related discussion). \RV{This can be seen to be conformally related to the ``classic'' Kerr-Schild form, as given in eq. 1.1 of \cite{Kerr+2009}.}

Therefore, for the AA metric itself to be asymptotically-flat, we require (a) $\lim_{r\rightarrow\infty}(2F/\Sigma) = 0$, and (b) $\lim_{r\rightarrow\infty}(X/\Sigma) = 1$. The first of these conditions is met due to the properties of the asymptotically-flat seed metric functions
\begin{align}
\lim_{r\rightarrow\infty}\frac{2F}{\Sigma} = \lim_{r\rightarrow\infty}\frac{(1/\sqrt{g} - f/g)R^2}{(f/g)R^2 + a^2\cos^2{\vartheta}} = 0\,, 
\end{align}
whereas the second is trivially met when $X = \Sigma$. Thus, asymptotically-flat seed metrics admit asymptotically-flat stationary generalizations when $X=\Sigma$. Note that this latter condition ($X=\Sigma$) is a sufficient but not a necessary condition. It is instructive to compare the asymptotically Minkowski metric in our siKS coordinates \eqref{eq:Asympt_Minkowski_siKS} with the analogous form of the Minkowski metric in the original (outgoing) Kerr-Schild coordinates for the Kerr metric, as given, e.g., in equations 1.7 and 1.13 of \cite{Wiltshire+2009}. 

\section{Separability of the geodesic equation for the Stationary metric}
\label{sec:AppC_Separability_AA_Metric}

The Lagrangian $\mathscr{L}$ describing a geodesic orbit $x^\mu(\lambda)$ is given as $2\mathscr{L} = u_\mu u^\mu$, where $u^\mu = \mathrm{d}x^\mu/\mathrm{d}\lambda = \dot{x}^\mu$ is the four-velocity along the geodesic and $\lambda$ is an affine parameter along it. Working now in Boyer-Lindquist coordinates \eqref{eq:AA_Metric}, due to the two Killing symmetries of the spacetime that are generated by $T = \partial_t$ and $\Phi = \partial_\varphi$, we can find momenta, $p_\mu = \partial_{\dot{x}^\mu}\mathscr{L} = \partial_{u^\mu}\mathscr{L} = u_\mu$, that are conserved, corresponding to the two cyclic variables. These are identified as being the energy $E = - u_\mu T^\mu$ and the azimuthal angular momentum $L = u_\mu \Phi^\mu$ of the orbit respectively,%
\footnote{For $K$ a Killing vector, $u^\mu\nabla_\mu(u^\nu K_\nu) = (u^\mu\nabla_\mu u^\nu)K_\nu + u^\mu u^\nu (\nabla_\mu K_\nu)  = 0$. The first and the second terms vanish due to the geodesic equation and the Killing equation respectively.} %
which can be used to obtain
\begin{align} \label{eq:tdot_phidot_Geodesic_Eqs}
\frac{\dot{t}}{E} &=\ -\frac{X}{\Sigma^2\det{[g_{t\varphi}]}}\left[\Pi - 2 F a \xi\right]\sin^2{\vartheta}\,, \\
\frac{\dot{\varphi}}{E} &=\ -\frac{X}{\Sigma^2\det{[g_{t\varphi}]}}\left[2 F a\sin^2{\vartheta} + (\Sigma - 2 F)\xi\right]\,, \nonumber
\end{align}
where we have introduced the first impact parameter, $\xi := L/E$, and the determinant of the $t\varphi-$sector of the AA metric tensor in BL coordinates \eqref{eq:AA_Metric}, $\det{[g_{t\varphi}]} := g_{tt}g_{\varphi\varphi} - g_{t\varphi}^2 \RV{= -(X^2/\Sigma^2)\Delta\sin^2{\vartheta}}$.


Now with eq. \ref{eq:tdot_phidot_Geodesic_Eqs} we can write,
\begin{align}
\frac{2\mathscr{L}X}{E^2}\Delta =&\
\left[X^2\frac{\dot{r}^2}{E^2} - \left(\Delta + 2F - a\xi\right)^2\right] \\
&\ + \Delta\left[X^2\frac{\dot{\vartheta}^2}{E^2} + \left(a\sin{\vartheta}-\xi\csc{\vartheta}\right)^2\right]\,. \nonumber
\end{align}
It is easy to see then that the geodesic equation for null geodesics ($2\mathscr{L} = 0$) is fully separable,
\begin{align}
X^2\frac{\dot{\vartheta}^2}{E^2} =&\ \eta^2 - \left(a\sin{\vartheta}-\xi\csc{\vartheta}\right)^2 =: \Theta_0(\vartheta)\,, \\
X^2\frac{\dot{r}^2}{E^2} =&\ \left(\Delta + 2F - a\xi\right)^2 - \Delta\eta^2 =: \mathscr{R}_0(r)\,,
\end{align}
where we have introduced a separation constant $\eta^2$ which is related to the Carter constant $C$ through $\eta^2 = C/E^2$. The Carter constant in turn can be used to demonstrate the existence of a Killing-Yano tensor and an associated hidden symmetry of the motion.

The fundamental principal null congruences (PNCs) of the AA spacetime consist of null geodesics that satisfy $\dot{\vartheta} = 0$ and $\ddot{\vartheta} = 0$ (see, e.g., \citealt{Misner+1973}, \citealt{Hioki+2008}; see also Sec. 2.3 of \citealt{Adamo+2009}). This is equivalent to requiring that $\Theta_0 = 0$ and $\partial_\vartheta\Theta_0 =0$, which yields a solution $\eta = 0$ and $\xi = a\sin^2{\vartheta}$. On a related note, the \cite{Newman+1962} complex null tetrad adapted to the outgoing ($\dot{r}>0$) PNC for the AA metric can be found in \cite{Azreg-Ainou2014a}.

If $X(r,\vartheta)$ is of the form $X(r,\vartheta) = X_r(r) + X_\vartheta(\vartheta)$, then the geodesic equation is separable even for non-null orbits,
\begin{align}
X^2\frac{\dot{\vartheta}^2}{E^2} =&\ \eta^2 - \left(a\sin{\vartheta}-\xi\csc{\vartheta}\right)^2 + \frac{2\mathscr{L}}{E^2}X_\vartheta =: \Theta(\vartheta)\,, \nonumber \\
X^2\frac{\dot{r}^2}{E^2} =&\ \left(\Delta + 2F - a\xi\right)^2 - \Delta\eta^2 + \frac{2\mathscr{L}}{E^2}\Delta X_r =: \mathscr{R}(r)\,.
\end{align}
For the special case when $X = \Sigma$ these are given simply as $X_r(r) = A(r)$ and $X_\vartheta(\vartheta) = a^2\cos^2{\vartheta}$.


\bibliography{Refs-Horizon-Penetrating-Coordinates-for-Axisymmetric-Spacetimes}

\end{document}